\documentclass[dvips,floatfix,intlimits,pra,showkeys,showpacs,twocolumn]{revtex4}

\usepackage{amsfonts,amsmath,amssymb}
\usepackage{dcolumn}
\usepackage{graphicx}
\usepackage{hypernat}
\usepackage{hyperref}
\usepackage{breakurl}

\hypersetup{%
   pdfauthor = {Joop~J.~Gilijamse, Jochen~K\"upper, Steven~Hoekstra,
      Sebastiaan~Y.~T.~van~de~Meerakker, Gerard~Meijer},%
   pdftitle = {Optimizing the Stark-decelerator beamline for the trapping of
      cold molecules using evolutionary strategies},%
   pdfsubject = {Published in Phys. Rev. A 73 (2006)},%
   pdfkeywords = {cold molecules; Stark deceleration; feedback control
      optimization; evolutionary strategies; machine learning}%
}

\begin{document}

\title{Optimizing the Stark-decelerator beamline for the trapping of cold
   molecules \\
   using evolutionary strategies}%
\author{Joop~J.~Gilijamse}%
\author{Jochen~K\"upper}%
\email[Author to whom correspondence should be addressed. Electronic address:
]{jochen@fhi-berlin.mpg.de}%
\author{Steven~Hoekstra}%
\author{Nicolas~Vanhaecke}%
\thanks{Present address: Laboratoire Aim\'e Cotton, CNRS II, Campus d'Orsay,
   B\^atiment 505, 91405 Orsay cedex, France.}%
\author{Sebastiaan~Y.~T.~van~de~Meerakker}%
\author{Gerard~Meijer}%
\affiliation{Fritz-Haber-Institut der Max-Planck-Gesellschaft, Faradayweg 4-6,
   14195 Berlin, Germany}%
\date{\today}%
\begin{abstract}\noindent%
   We demonstrate feedback control optimization for the Stark deceleration and
   trapping of neutral polar molecules using evolutionary strategies. In a
   Stark-decelerator beamline, pulsed electric fields are used to decelerate OH
   radicals and subsequently store them in an electrostatic trap. The efficiency
   of the deceleration and trapping process is determined by the exact timings
   of the applied electric field pulses. Automated optimization of these timings
   yields an increase of 40~\% of the number of trapped OH radicals. \\

   \noindent Copyright 2006 American Physical Society. This article may be
   downloaded for personal use only. Any other use requires prior permission of
   the author and the American Physical Society. The following article appeared
   in \emph{Phys. Rev.} A \textbf{73}, 063410 (2006) and may be found at
   \url{http://dx.doi.org/10.1103/PhysRevA.73.063410}.
\end{abstract}%
\keywords{cold molecules; Stark deceleration; feedback control optimization;
   evolutionary strategies; machine learning}%
\pacs{33.80.Ps, 39.10.+j, 02.70.-c}
\maketitle%

\section{Introduction}
\label{sec:introduction}

Analogous to the interaction of charged particles with electric fields in a
linear accelerator \cite{Lee:AccPhys:2004}, the interaction of neutral polar
molecules with electric field gradients can be used in a Stark decelerator
\cite{Bethlem:PRL83:1558} to accelerate, decelerate, or guide a molecular beam.
Using arrays of electric field electrodes that are switched to high voltage at
appropriate times, bunches of state-selected molecules with a
computer-controlled velocity and with a low longitudinal temperature can be
produced. This is of advantage in any molecular beam experiment where the
velocity distribution of the molecules is an important parameter. When combined
with an electrostatic trap, the Stark-deceleration technique offers the
possibility to confine rovibronic ground-state molecules for times up to
seconds~\cite{Bethlem:Nature406:491,Meerakker:PRL94:023004}. This holds great
promise for the study of molecular interactions at the high densities and the
(ultra) low temperatures that can ultimately be achieved~\cite{EPJD31:ColdMol}.

The efficiency of the deceleration and trap-loading process critically depends
on the exact timings of the high-voltage pulses. In a typical deceleration and
trapping experiment a sequence of more than 100 high-voltage pulses is applied
to the various elements in the beamline. The time sequence that is used is
inferred from a detailed knowledge of the electric fields in the decelerator and
trap region, and the Stark effect of the molecule of interest. This, however,
does not account for possible deviations from an idealized description of the
experiment, such as, for instance, misalignments of the electrode arrays and
instabilities of the applied high-voltage pulses. Furthermore, these
calculations are based on a one-dimensional model to describe the longitudinal
motion, while the transverse motion of the molecule effects the efficiency of
the decelerator \cite{Meerakker:PRA73:023401}. A manual optimization of the time
sequence is practically impossible for this complicated and large parameter
space. Here, we demonstrate the successful implementation of an evolutionary
algorithm for the automated optimization of a Stark-decelerator beamline.

Evolutionary algorithms (EA), mimicking the biological principles of evolution,
have been frequently used for automatic optimization of experimental problems
with a large parameter space and noisy feedback signals. As early as the 1960s,
three independent developments started with the introduction of evolutionary
strategies (ES) by Rechenberg and Schwefel
\cite{Rechenberg:WGLR:1964,Rechenberg:Dissertation:1971,Schwefel:EvolOptSeek},
evolutionary programming (EP) by Fogel, Owens, and Walsh
\cite{Fogel:AISimEvol:1965,Fogel:AISimEvol:1966}, and genetic algorithms (GA) by
Holland \cite{Holland:Adaption:1975,Goldberg:GeneticAlgorithms:2002}. A nice
introduction to the field of evolutionary computing and its different dialects
is given by Eiben and Smith \cite{Eiben:IntroEvolComp}.

In many branches of atomic and molecular physics feedback control experiments
have been performed; see, for example reference~\onlinecite[and references
therein]{Rice:OpticalControlMolDyn}. Since the proposal \cite{Judson:PRL68:1500}
and application \cite{Assion:Science282:919} of learning loops to optimize
femtosecond laser pulse shapes \cite{Goswami:PhysRep374:385} for the control of
quantum dynamics in the 1990s, a large number of experiments on the coherent
control of atomic and molecular dynamics have been performed
\cite{Brixner:AAMOP46:1,Levis:JPCA106:6427,Brixner:CPC4:418}.

In this work, we use evolutionary strategies for the feedback control
optimization of the time sequence of high-voltage pulses that are applied to the
Stark decelerator and trap. The experiments have been performed using a pulsed
molecular beam of OH radicals in the low-field seeking
$X\,^2\Pi_{3/2},v=0,J=3/2,M\Omega=-9/4$ state, for which Stark deceleration
\cite{Bochinski:PRL91:243001} and electrostatic trapping
\cite{Meerakker:PRL94:023004} had previously been demonstrated. The automated
optimization results in an increase of up to 40~\% of the number of trapped OH
radicals.

\section{Experimental setup}
\label{sec:experimental}

\subsection{Stark deceleration and trapping}
\label{sec:experimental:deceleration}

Molecules possessing an electric dipole moment will gain Stark energy upon
entering an electric field, when in an appropriate quantum state. This gain in
Stark energy is compensated by a loss in kinetic energy. If the electric field
is switched off before the molecules have left the field, they will not regain
the lost kinetic energy. In a Stark decelerator
\cite{Bethlem:PRL83:1558,Bethlem:PRA65:053416}, this process is repeated by
letting the molecules pass through multiple switchable electric field stages. In
this way, molecules can be decelerated and brought to a standstill.

\begin{figure}
   \centering
   \includegraphics[width=\linewidth]{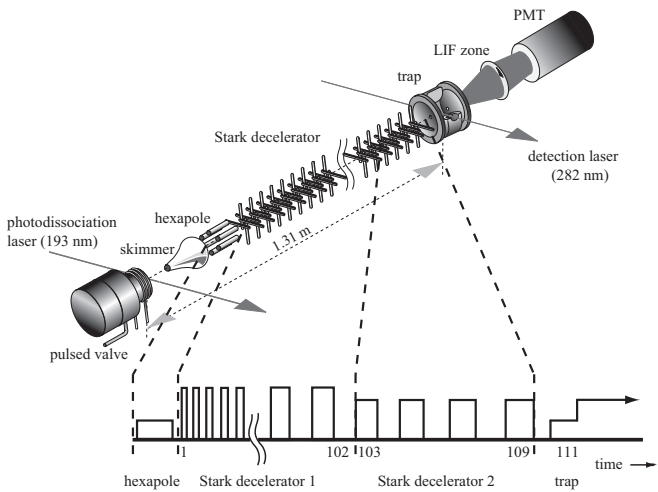}
   \caption{Scheme of the experimental setup. A pulsed beam of OH radicals with
      a mean velocity of 360~m/s is produced via ArF-laser photodissociation of
      HNO$_3$ seeded in Xe. The molecules pass through a skimmer, a hexapole,
      and a Stark decelerator and are subsequently confined in an electrostatic
      trap. State-selective LIF detection is performed inside the trap. A
      schematic representation of the time sequence of high-voltage pulses is
      shown at the bottom of the figure, including selected indices of the
      switching times; see text for details.}
   \label{fig:setup}
\end{figure}

The experimental setup is schematically shown in Fig.~\ref{fig:setup}, and is
described in detail elsewhere \cite{Meerakker:ARPC57:159}.
In brief, a pulsed beam of OH radicals is produced by photodissociation of
HNO$_3$ that is co-expanded with Xe from a pulsed solenoid valve. The mean
velocity of the beam is around 360~m/s with a velocity spread (full width at
half maximum) of 15~\%. After the supersonic expansion, most of the OH radicals
in the beam reside in the lowest rotational ($J=3/2$) level in the vibrational
and electronic ground state $X\,^2\Pi_{3/2},v=0$. The molecular beam passes
through a skimmer with a 2~mm-diameter opening and is transversely focused into
the Stark decelerator using a short pulsed hexapole. The Stark decelerator
consists of an array of 109 equidistant pairs of electrodes, with a
center-to-center distance of 11~mm. The decelerator is operated using a voltage
difference of 40~kV between opposing electrodes, creating a maximum electric
field strength on the molecular beam axis of about 90~kV/cm. A kinetic energy of
0.9~cm$^{-1}$ is extracted from the OH molecules per deceleration stage (the
region between adjacent pairs of electrodes), and part of the beam is
decelerated from 371 to 79~m/s after 101 stages. In the remainder of this paper,
these first 101 stages will be referred to as decelerator~1. The last seven
stages of the decelerator, referred to as decelerator~2, are electronically and
mechanically decoupled from decelerator~1, and are used at a lower voltage
difference of $30$~kV. Here, the molecules are decelerated further to a velocity
of $21$~m/s, prior to the loading of the packet into the electrostatic trap. The
trap consists of two hyperbolic endcaps and a ring electrode. To load the
molecules into the trap its electrodes are switched from an initial loading
configuration to a trapping configuration. The loading configuration creates a
potential hill that is higher than the kinetic energy of the molecules. The OH
radicals, therefore, come to a standstill while flying into the trap. At this
moment the electrodes are switched to the trapping configuration, creating a
field minimum in the center of the trap.

The number of trapped OH radicals as well as the temperature of the trapped gas
critically depend on the details of the trap-loading sequence, and in particular
on the velocity with which the molecules enter the trap
\cite{Meerakker:PRL94:023004}. If this velocity is chosen such that the
molecules come to a standstill exactly at the center of the trap (v$=15$~m/s), a
distribution corresponding to a temperature of $50$~mK can be reached. If this
velocity is larger, the molecules come to a standstill past the center of the
trap, and the final temperature is higher. The reduced spreading out of a faster
beam while flying from the last stage of the decelerator to the trap, however,
results in a larger number of trapped molecules. The velocity of 21~m/s and the
subsequent trap-loading sequence that is used as reference for the optimization
in the present experiment are identical to the trap-loading that was used in
previous OH trapping experiments \cite{Meerakker:PRL94:023004}. It results in a
temperature of the trapped molecular packet of about 450~mK, an estimated number
density of $10^{7}-10^{8}$ molecules per cm$^{3}$, and a trapping lifetime of
$1.6$~s.

The OH radicals are state selectively detected in the trap using a laser-induced
fluorescence (LIF) detection scheme. The 282~nm UV radiation of a pulsed dye
laser excites the $A\,^2\Sigma^+,v=1\longleftarrow{}X\,^2\Pi_{3/2},v=0$
transition. A photomultiplier tube (PMT) is used to measure the resulting
off-resonant fluorescence. In the experiments reported here, the repetition rate
of the experiment is 10~Hz and for every datapoint 64 successive measurements
are averaged. The signal-to-noise ratio of the trapping experiment under these
conditions is about 20.

\subsection{Feedback control optimization}
\label{sec:experimental:control}

As described in section~\ref{sec:experimental:deceleration} the individual
timings in the time sequences applied to the machine are very critical.
Generally, initial time sequences are calculated based on a theoretical model of
the experiment and will be referred to as \emph{calculated} time sequences
throughout this paper.

For the feedback control optimization the LIF intensity of trapped OH molecules,
as described above, is used. To avoid effects from the oscillations of the
molecular packet inside the trap that appear during the first milliseconds after
switching on the trap (see Fig.~3 of \cite{Meerakker:PRL94:023004} and
Fig.~\ref{fig:tof} of this paper), the LIF intensity is measured after $20$~ms
trapping-time. This measurement of the OH density in the trap is used as
objective function (fitness) in the feedback control algorithm. Since the
lifetime of the OH radicals confined in the trap is as long as $1.6$~s, the
number of detected OH molecules after 20~ms is still $>98$~\% of the maximum
value. Because the LIF signal at that detection time is practically constant
over periods much longer than the timing changes due to the feedback control
algorithm ($\ll{}1$~ms), pulsed laser excitation at a fixed time can be applied
for the molecule detection. Note that in the feedback control loop implemented
here, we use the result from previous experimental runs as feedback for
following ones.

This given problem requires the optimization in a large parameter space, which
at the same time can only be sampled by a slow and noisy evaluation. For such
problems evolutionary algorithms are generally a good choice and have been
applied successfully in many fields. The individual parameters to be adjusted
are the timings $t_i$ that determine the exact switching of the high voltages
energizing the deceleration and trapping electrodes. For the given experiment
this results in 111 parameters to be optimized. For a detailed depiction of the
timing numbering see Fig.~\ref{fig:setup}. To reduce the high dimensionality of
the parameter space, we retracted from optimizing all parameters individually,
but encoded them in three reduced sets of parameters: The timings of
decelerator~1 and the first four timings of decelerator~2 are not optimized
independently, but described by two sets of polynomial expansion coefficients.
We found that an accurate encoding of the time sequence itself requires a
polynomial of high order, i.\,e., orders larger than 20 for a 5~$\mu$s accuracy.
To allow for smaller polynomial orders $o_1$ and $o_2$ for the two parts of the
decelerator, we have only encoded the differences to the calculated time
sequence $t_i-t_{i,0}$ in the polynomial, allowing for considerably smaller
expansions, since they only need to describe deviations from the theoretical
timings. For decelerator~1, one obtains timings $t_i$ with $i=1$--$102$
\begin{align}
   t_i = t_{i,0} + \sum_{j=0}^{o_\text{1}} p_{j+1} \cdot (i-1)^j
\end{align}
and for decelerator~2 timings $t_i$ with $i=103$--$106$
\begin{align}
   t_i = t_{i,0} + \sum_{j=0}^{o_\text{2}} p_{j+o_1+2} \cdot (i-103)^j
\end{align}
The remaining five timings $t_i$ for the last deceleration stages and the
trap-loading and trapping configurations, which are the most critical timings,
are optimized individually and independently. To decouple them from the changes
of earlier timings, they are encoded as time difference to their respective
preceding timing, i.\,e., we use
\begin{equation}
   \label{eq:es:delta-t}
   \Delta{}t_i = t_i - t_{i-1} = p_{i+o_1+o_2-104} 
\end{equation}
for $i=107$--$111$. The complete parameter vector used in the optimization
is then encoded as
\begin{equation}
   \vec{P} = { \left(p_1,p_2,\ldots{}p_{o_1+o_2+7}\right)^T
      \in \left(\mathbb{R}^+\right)^{o_1+o_2+7} }
\end{equation}
Typically we have used polynomials of order \mbox{$o_1=2$} for decelerator~1 and
order \mbox{$o_2=1$} for decelerator~2, resulting in a parameter vector of
length ten. In this way, the dimension of the parameter space is reduced by one
order of magnitude compared to the initial one, while control over the whole
beamline by the feedback loop is maintained.

With the intuitive representation of the individuals of the optimization problem
as a vector of real numbers over a continuous parameter space, the choice of
evolutionary strategies is a natural one. ES is an EA dialect that uses a
representation of real-valued vectors and generally uses self-adaptivity
\cite{Eiben:IntroEvolComp}. In the experiments described here, we used the
Evolving Object (EO) framework \cite{Keijzer:ArtEvol:2310:231,EO:web}
implementation of the ES. As a trade-off between problem size in the ES and
theoretical convergence, the \emph{eoEsStdev} ES strategy, applying uncorrelated
mutations with individual step sizes, was used
\cite[section~4.4.2]{Eiben:IntroEvolComp}. In this self-adaptive strategy the
genotype is a vector of real numbers containing the actual optimization
parameters as well as individual mutation widths $\sigma_i$ for every parameter
$p_i$.

The initial optimization meta-parameters used were based on the suggestions by
Eiben and Smith \cite{Eiben:IntroEvolComp} and successively adopted according to
their success in the experiments. In the most successful optimization runs the
following parameters were used: typically a population size of five or ten
individuals was used, with population sizes up to 40 in some runs. Typically 30
offsprings were generated every generation, with values ranging from the actual
population size to six times the population size over different runs. Generally,
an offspring-to-population ratio of seven is assumed to work best, but the
theoretical advantage is apparently outweighed by the slowness of the evaluation
and the corresponding experimental difficulties in this experiment. The most
successful mutation and crossover rates were 75~\% and 50~\%, respectively, but
this seems not to be critical and was not tested extensively. Parent selection
was done using the ordered sequential selector. We have used discrete global
recombination for the experimental parameters and intermediate global
recombination for the mutation widths $\sigma$. For survivor selection the
$(\lambda,\mu)$ approach worked best, as it seems to handle noise and drifts in
the experimental conditions well, as is generally assumed
\cite[section~4.7]{Eiben:IntroEvolComp}. Elitism was not applied.

This machine learning is implemented in our data-acquisition system (KouDA)
using ES within an automatic feedback control loop.

\section{Experimental results}
\label{sec:results}

\begin{figure}
   \centering
   \includegraphics[width=\linewidth]{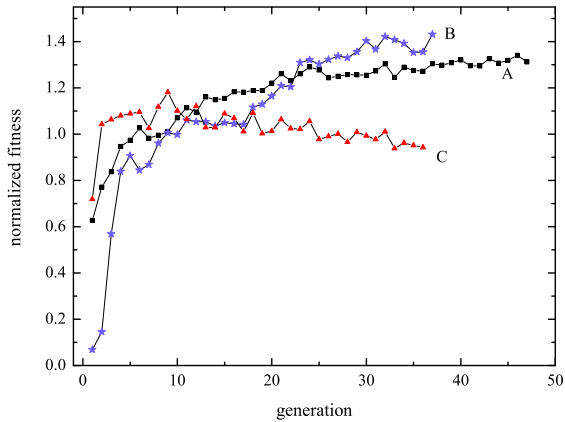}
   \caption{(Color Online): For three different optimization runs, each with
      different initial parameters, the normalized average fitness per
      generation is plotted. Curves~A (squares) and B (asterisks) show an
      increase of 30 and 40~\%, respectively. During the measurement represented
      by curve~C (triangles) drifts in the experimental conditions, namely the
      backing pressure of the supersonic expansion, occurred and led to reduced
      signal intensities, as was confirmed after the optimization run.}
   \label{fig:bestavg}
\end{figure}

In Fig.~\ref{fig:bestavg} the normalized average fitness --- the LIF signal from
OH radicals in the trap --- per generation is plotted against the generation
number for three different optimization runs, referred to as runs A, B, and C.
The measured fitness-values are normalized with respect to the fitness obtained
for the calculated time sequence under the same experimental conditions. In each
run, different strategy parameters for the algorithm or different initial
populations are used, as detailed below. Typically, a complete optimization run
corresponds to the evaluation of many hundred generated time sequences and takes
about 1--2~h of measuring time.
In run A (squares), the calculated time sequence is used as starting point for
the optimization. From this sequence, an initial population is created with
parameters that are randomly picked out of a Gaussian distribution around the
calculated values. The last parameter, $\Delta{}t_{111}$, has been decreased by
27~$\mu$s based on the outcome of earlier runs (not presented). As a result of
these small changes, the first generation has a slightly lower fitness. After
nine generations the average fitness of the generation has increased to the
value of the calculated time sequence. For later generation numbers the fitness
increases further and reaches a maximum of 1.3 after 46 generations. In the
measurement represented by curve~B (asterisks) an initial population was created
from the same calculated time sequence, but nine out of ten parameters were set
off by 3 to 20~\%. Hence, the first generation time sequences lead to a
normalized fitness of less than 0.1. After 11 generations this number already
reaches 1 and is further optimized to 1.4 in generation 37. The optimization
runs~A and B result in a number of trapped OH radicals that is 30 to 40~\%
higher than the number that is obtained with the calculated time sequence. Other
experiments in which different initial populations were chosen led to a similar
increase in the number of trapped molecules.

The initial population and strategy parameters, which are used in the
optimization run shown in curve~C (triangles), are very similar to the
parameters that were used in curve~A. Curve~C initially shows (as expected) an
optimization similar to that of run A and reaches a maximum of 1.2 after around
nine generations. From then on, however, the fitness starts decreasing. This is
due to a drift in the production of OH radicals during this experimental run,
that was confirmed by an independent measurement after the optimization run. In
spite of this drift the algorithm still converged and the time sequences
obtained for the last generation are comparable with time sequences obtained in
runs A and B (\textit{vide infra}).

Other experiments using different strategy parameters for the ES, for example,
different population sizes or different settings for mutation and crossover
rates, did lead to a similar increase in the number of trapped molecules of
35--40~\%. Furthermore, the values of corresponding parameters from the
optimized time sequences are generally comparable. These results show not only
that the algorithm is able to optimize the number of trapped molecules, but also
that it finds a reproducible maximum in the parameter-space, even if the initial
parameters deviate significantly or external factors disturb the experiment.

\begin{figure}
   \centering
   \includegraphics[width=\linewidth]{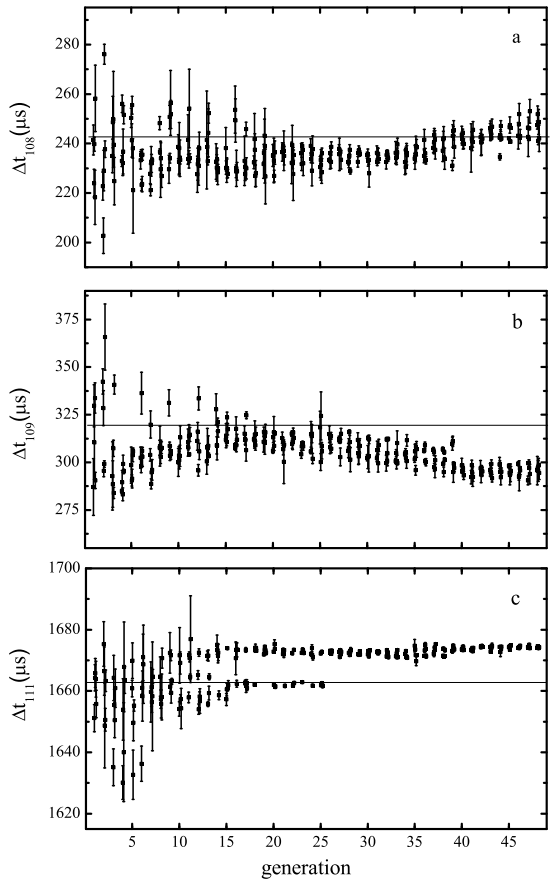}
   \caption{The evolution of three of the parameters during optimization run~A
      (see Fig.~\ref{fig:bestavg}): a)~$\Delta{}t_{108}$, b)~$\Delta{}t_{109}$,
      c)~$\Delta{}t_{111}$. The squares mark the value of individual parameters
      and the error bars represent the corresponding mutation widths $\sigma$.
      Only the five parameters selected for a new population (as described in
      section~\ref{sec:experimental:control}) are shown and they are grouped by
      generation. Individual parameters within a population are slightly offset
      horizontally to allow the observation of individual values and their error
      bars. At the beginning, a large range of the parameter space is searched,
      whereas later in the optimization the $\sigma$'s are reduced by the
      algorithm and convergence is reached. The horizontal lines denote the mean
      value of each parameter in the first generation.}
   \label{fig:genes}
\end{figure}

The evolutions of three of the most important parameters, recorded during
optimization run A, are shown in Fig.~\ref{fig:genes}. Fig.~\ref{fig:genes}\,a
and \ref{fig:genes}\,b show $\Delta{}t_{108}$ and $\Delta{}t_{109}$,
respectively. These parameters define the switching times of the last two stages
of decelerator~2 and thus determine the exact velocity with which the molecules
leave the decelerator. Fig.~\ref{fig:genes}\,c depicts the evolution of
$\Delta{}t_{111}$, the time interval during which the loading configuration of
the trap is used. At the end of this time interval the trapping configuration is
switched on. For reference, the horizontal lines in the plots denote the mean
value of the respective parameter in the first generation, which are equivalent
to the parameters in the calculated time sequence. Although the fitness depends
very critically on these specific timings, the evolution of the parameters shown
in Fig.~\ref{fig:genes} is typical for the evolution of less critical parameters
as well.

For all three parameters, the mutation widths $\sigma$, represented by the
vertical bars, are initially large and the parameters scatter over a relatively
large range. As the generation number increases, this mutation width decreases
and the parameters converge. Parameter $\Delta{}t_{111}$, however, converges
initially to two values, one centered around 1662~$\mu$s, the other around
1674~$\mu$s. This shows that the parameter-space contains multiple local maxima,
and that multiple pathways in the parameter-space can be followed. Only after 27
generations, exclusively individuals with a value for $\Delta{}t_{111}$ of about
1674~$\mu$s survive the selection. As most runs converge to similar values this
seems to be the global optimum, at least for the parameter space searched.

From each feedback control experiment a set of optimized time sequences is
obtained. It is clear from the optimized time sequences, that no different mode
of operation for the Stark decelerator is obtained and that the previous
theoretical understanding \cite{Bethlem:PRA65:053416} is confirmed by these
experiments. Moreover, comparing the time-of-flight (TOF) profiles of OH
radicals at the center of the trap, which are measured using the calculated and
optimized time sequences, a physical interpretation of the differences can be
deduced.
\begin{figure}
   \centering
   \includegraphics[width=\linewidth]{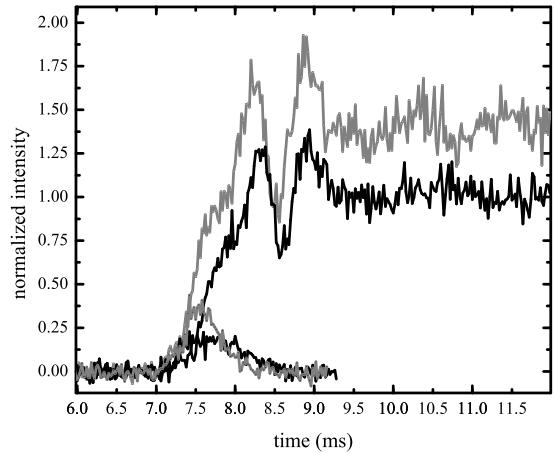}
   \caption{Density of OH radicals at the center of the trap as a function of
      time after their production. The lower two traces are the intensities of
      molecules passing through the center of the trap without any voltages
      applied to the trap electrodes. The upper traces are measurements for
      trapping experiments. The black traces are measured with the calculated
      time sequence applied to the machine, whereas the gray traces are measured
      with one of the generated, optimized time sequences obtained from
      automated optimization using evolutionary strategies.}
   \label{fig:tof}
\end{figure}
The typical result of such a measurement is shown in Fig.~\ref{fig:tof}. The
black and gray curves are measured using the calculated and optimized
time sequences, respectively. The lower two curves show the TOF profiles of the
OH molecules as they arrive in the trap when no voltages are applied to the trap
electrodes. The positions and widths of the arrival-time distributions are a
measure for the longitudinal velocity distributions of the decelerated OH beams
that exit the decelerator. Compared to the calculated time sequence, the
optimized sequence results in an arrival-time distribution that is shifted
180~$\mu$s to the left, indicating that the molecular packet arrives with a
higher mean velocity of 25~m/s, instead of 21~m/s, in the trap. Assuming the
transverse and longitudinal velocity spreads are unaltered for the optimized
time sequence, the beam spreads out less in all directions while traveling the
distance from the end of the decelerator to the trap, and the corresponding
arrival-time distribution is narrower. The integral of the peak of the arriving
packet (lower curves) is already enhanced by about 40~\%, reflecting the reduced
transverse spreading out of the beam and hence the reduced transverse losses
while entering the trap.

The upper two curves show the density of OH radicals at the center of the trap
when the trap-loading and trapping electric fields are applied. The optimized
time sequence (gray curve) leads to a more pronounced oscillation in the TOF
profile than the calculated one (black curve). This is readily understood from
the higher initial velocity of the molecules. The molecules enter the trap too
fast, and come to a standstill past the center of the trap. The molecular packet
is poorly matched to the trap acceptance, and the width of the velocity
distribution of the trapped molecules will therefore be larger as well. These
results confirm, as was already concluded earlier~\cite{Meerakker:PRL94:023004},
that a large number of molecules in the trap and a low temperature of the
trapped packet of molecules are conflicting goals with the present design of the
trap: the required low velocity to match the decelerated molecular packet with
the acceptance of the trap results in a large transverse spreading out of the
packet prior to entering the trap.

In principle, one could also aim a feedback control optimization at determining
a time sequence for a trapped molecular packet with a temperature as low as
possible, or a weighted combination of the number of trapped molecules and a
minimal temperature, by using an appropriate experimental objective function.
One could, for example, measure the number of molecules at the center of the
trapping region after a predefined time of free expansion of a previously
trapped packet. That would result in a combined determination of the peak
density of the trapped molecular packet and its temperature, where the time
delay between switching off the trap and the detection of the molecular density
would weigh the two contributions to the fitness. Alternatively, if the spatial
density distribution of the trapped molecular packet would be measured for every
generated time sequence, direct information on the number of trapped molecules
and their temperature is obtained, allowing to define any objective function
based on these two important measures. Furthermore, when using continuous
detection to allow for measuring the complete time-of-flight profile from the
nozzle to the detection region for every molecular packet, the integrated
intensity and the longitudinal temperature can be deduced offline by the
optimization algorithm~\footnote{The cw detection of molecular packets in a
   Stark-decelerator beamline has been demonstrated for CO ($a\,^3\Pi$)
   \cite{Bethlem:PRL83:1558}, YbF \cite{Tarbutt:PRL92:173002}, OH, and
   benzonitrile (our laboratory).}. This allows to optimize any
Stark-decelerator beamline, even without trapping.

Besides the timings of the high-voltage pulses one can also optimize other
computer controllable experimental parameters, such as the voltages that are
applied to the experiment, laser frequencies, etc. In general, evolutionary
algorithms can be used for the optimization of any fitness function that can be
determined experimentally. This includes, for example, the ratio of molecules
simultaneously trapped in two different quantum states or the ratio of
decelerated and actually trapped molecules. More generally, the method can also
be applied to other atomic and molecular beam experiments, such as optimizing
the timings or voltages in multipole focusers~\cite{Reuss:StateSelection} or the
currents in a Zeeman slower~\cite{Metcalf:LaserCooling}.\\

\section{Conclusions}
\label{sec:conclusions}

In this paper we describe the successful implementation of feedback control
optimization of the Stark deceleration and trapping of OH radicals using
evolutionary strategies. The time sequence of high-voltage pulses that is applied
to the decelerator and trap electrodes is encoded as parameter vector for the
algorithm. Starting from an initial time sequence based on an idealized
representation of the beamline, the number of trapped OH radicals is increased
by 40~\%. This enhancement is qualitatively understood in terms of the improved
coupling in of the amount of molecules into the trap.

The machine learning approach presented here can be applied to other
Stark-deceleration experiments as well. The optimization will be especially
useful for all experiments in which very slow molecular beams ($v<100$~m/s) are
manipulated, for which the exact switching times of the high-voltage pulses are
extremely critical. In general, any computer-controllable experimental parameter
can be optimized using evolutionary algorithms and any fitness function that can
be determined experimentally can be used as fitness for the optimization.

Essential to the present experiment is the use of trapped molecules, which
enables the decoupling of the timing for pulsed laser detection from the
optimization. For beamlines with continuous detection such a timing can be
evaluated offline and becomes uncritical, thus making feedback control
optimization generally applicable.

\begin{acknowledgments}
   This work is supported by the European Union Research and Training Network
   ``Cold Molecules''.
\end{acknowledgments}
\bibliography{string,mp}

\end{document}